\documentclass[]{spie}  

\newcommand{\JCMsuite}{\textsc{JCMsuite}\cite{jcmsuite},}

\newcommand{\bbz}{\mathbb{Z}}
\newcommand{\bbn}{\mathbb{N}}
\newcommand{\bbr}{\mathbb{R}}

\newcommand{\ud}{\,\mathrm{d}}

\usepackage{amsmath,amsfonts,amssymb}
\usepackage{graphicx}
\usepackage[colorlinks=true, allcolors=blue]{hyperref}

\title{Forward simulation of coherent beams on grating structures for coherent scatterometry}

\author[a,b]{
  Martin~Hammerschmidt}
\author[a,b]{Lin~Zschiedrich}  
\author[c]{Lauryna Siaudinyté}  
\author[a,b]{Phillip Manley}  
\author[a,b]{Philipp-Immanuel Schneider} 
\author[a,b]{Sven~Burger}  
\affil[a]{
JCMwave GmbH\\
Bolivarallee~22, 
D\,--\,14050 Berlin,
Germany}
\affil[b]{Zuse Institute Berlin\\
Takustra{\ss}e~7,
D\,--\,14195 Berlin,
Germany}
\affil[c]{VSL, National Metrology Institute \\Thijsseweg 11\\
2629 JA Delft\\
Netherlands}

\authorinfo{Further author information: (Send correspondence to M.~Hammerschmidt)\\M.~Hammerschmidt: E-mail: martin.hammerschmidt@jcmwave.com, 
URL: http://www.jcmwave.com\\
URL: http://www.zib.de}

\begin{document} 
\maketitle

\begin{abstract}
Modelling the scattering of focused, coherent light by nano-scale structures is oftentimes used to reconstruct or infer geometrical or material properties of structures under investigation in optical scatterometry. This comprises both periodic and aperiodic nano-structures. Coherent Fourier scatterometry with focused light exploits the diffraction pattern formed by the nano-structures in Fourier plane.  While the scattering of a focused beam by a spatially isolated scatterer is a standard modelling task for state-of-the art electromagnetic solvers based, e.g., on the finite element method, the case of periodically structured samples is more involved. In particular when the focused light covers several grating periods of as it is commonly the case. 

We will present a coherent illumination model for scattering of focused beams such as Gaussian- and Bessel- beams by periodic structures such as line gratings. The model allows to take into account optical wavefront aberrations in optical systems used for both, the illumination and detection of the scattered fields. We compare the model with strategies implemented on large-scale super-cells and inverse Floquet-transform strategies to superimpose both near- and far fields coherently.
\end{abstract}


\textit{This paper will be published in Proc. SPIE PC12619, Modeling Aspects in Optical Metrology IX, PC1261907, DOI: https://doi.org/10.1117/12.2673231) and is made available as an electronic preprint with permission of SPIE. One print or electronic copy may be made for personal use only. Systematic or multiple reproduction, distribution to multiple locations via electronic or other means, duplication of any material in this paper for a fee or for commercial purposes, or modification of the content of the paper are prohibited.}

\section{INTRODUCTION}
\label{sec:intro}  
Scatterometry is an inverse measurement technique which reconstructs geometrical and/or material properties of structures under investigation from a parameterized forward model of the measurement process. The success and validity of the reconstruction framework relies heavily on efficient numerical tools to solve Maxwell's equations to simulate the measurement process for a given parameter configuration.

State-of-the art electromagnetic solvers whether they are based on the finite element method\cite{Pomplun2007} (FEM), finite-difference time-domain method or rigorous coupled wave analysis, can be employed to model these scattering processes in the measurement. Structured samples require generally more involved simulations than planar or thin-film geometries encountered in ellipsometry where semi-analytical methods (such as the transfer matrix method) can be employed. In these cases a more general (and thus complex) method is required to model the geometry as accurately as required for the desired simulation accuracy.   

Coherent Fourier scatterometry\cite{kumar2014reconstruction,Siaudinyte_2020} (CFS) promises higher sensitivity among the currently used scatterometry techniques due to the use of coherent illumination instead of the commonly used incoherent illumination. In CFS measurement the sample is illuminated with a focused coherent beam of polarised, monochromatic illumination (typically through a microscope objective), then the scattered light is collected and the far-field is analysed in the back-focal plane of the    microscope objective. The technology can be used to analyse both periodic and aperiodic samples. 

Modelling the scattering of a focused beam by a spatially isolated scatterer is a standard modelling task for state-of-the art electromagnetic solvers. The case of periodic nano-structures is more involved as the focused beam spot covers several grating periods. Modelling the finite sized spot with strategies implemented on large-scale super-cells is straight forward but comes at significant computational costs which may prohibit a parameter reconstruction at reasonable computational costs through solving an optimization problem.

We will thus focus on techniques to restrict computations to the unit cell of periodic structures such as line gratings. We present the Inverse Floquet Transform strategy in Section \ref{sec:invFloquetTrafo} to superimpose near fields coherently and apply it to an example. In Section \ref{sec:imaging} we present a coherent illumination model for scattering of coherent focused beams, such as Gaussian and Bessel beams, by periodic structures. The model allows to take into account optical wavefront aberrations in optical systems used for both the illumination and detection of the scattered fields. Finally we apply it to a CFS example in Section~\ref{sec:cfs_example}.

\section{Inverse Floquet Transform to model focused light scattered by periodic gratings}
In this section we present and apply the Inverse Floquet Transform to model the interaction of a coherent Gaussian beam with a grating in the nearfield. 

\subsection{Inverse Floquet Transform Method}
\label{sec:invFloquetTrafo}
The Inverse Floquet Transform can be used to model the effects of an aperiodic (or isolated) source in a periodic structure. The approach has been applied before  to model scattering of a dipole in a periodic environment~\cite{ZschiedrichGreinerBurgeretal.2013}. 

Here, we restrict the discussion to periodic setups in 2D. This means that the following relation of the material tensors with respect to the lattice vector $\mathbf{a} = \begin{pmatrix}a\\0\end{pmatrix}$ holds: 
\begin{align*}
    \varepsilon(\mathbf{r}+\mathbf{a}) &=   \varepsilon(\mathbf{r}), \\ 
    \mu(\mathbf{r}+\mathbf{a}) &=   \mu(\mathbf{r}).
\end{align*}
A field is said to be Bloch-periodic if it satisfies the following condition for all $m\in \bbz$ with respect to a given Bloch vector $\mathbf{k_B} \in \bbr^2$:
\begin{equation*}
    \mathbf{E}(\mathbf{r}+m\mathbf{a}) = \mathbf{E}(\mathbf{r})e^{i\mathbf{k_B}^T\mathbf{a}m}
\end{equation*}
This relation allows to define Bloch-periodic boundary conditions and restrict calculations in periodic setups to the unit cell.

The Floquet transform of a field $\mathbf{E}(\mathbf{r})$  for a given  Bloch vector $\mathbf{k_B}$ can be calculated by 
\begin{align*}
    \mathbf{E}_{\mathbf{k_B}}(\mathbf{r}) = \sum_{m\in \bbn} \mathbf{E}(\mathbf{r}-m\mathbf{a})e^{i\mathbf{k_B}^T\mathbf{a}m},
\end{align*}
provided $\mathbf{E}$ is sufficiently localized and decaying.  $\mathbf{E}_{\mathbf{k_B}}(\mathbf{r}) $ is obviously Bloch-periodic by construction and can thus be calculated on the unit cell. 

Computing $\mathbf{E}_{\mathbf{k_B}}(\mathbf{r}) $ for all Bloch vectors $\mathbf{k_B}$ is not necessary as it suffices to compute the Floquet transform for vectors within the first Brillouin zone. 

The Brillouin zone in a one-fold periodic domain is given by 
\begin{align*}
    BZ = \left\{ \mathbf{k_B} = \mathbf{b}\tau \, \vert \, \tau \in [0,1] \right\}
\end{align*}
where the reciprocal lattice vector $\mathbf{b} = \begin{pmatrix}\frac{2\pi}{a}\\0\end{pmatrix}$ is related to the lattice periodicity $a$.

Integrating $\mathbf{E}_{\mathbf{k_B}}(\mathbf{r}) $ over the first Brillouin zone $BZ$ in fact recovers the original field $\mathbf{E}$:
\begin{align*}
    \int_{BZ} \mathbf{E}_{\mathbf{k_B}}(\mathbf{r}) \ud \mathbf{k_B} &= \int_{BZ} \sum_{m\in \bbn} \mathbf{E}(\mathbf{r}-m\mathbf{a})e^{i\mathbf{k_B}^T\mathbf{a}m}\ud \mathbf{k_B} \\ 
     &= \frac{2\pi}{a} \sum_{m\in \bbn}  \mathbf{E}(\mathbf{r}-m\mathbf{a}) \underbrace{\int_{0}^{1}e^{i\tau\mathbf{b}^T\mathbf{a}m}\ud \tau}_{=\delta_{m,0} \, \text{as} \, \mathbf{b}^T\mathbf{a} = 2\pi} \\ 
     & = \frac{2\pi}{a} \mathbf{E}(\mathbf{r}).
\end{align*}

In practice we will approximate the integral over the Brillouin zone $BZ$ by a quadrature algorithm and refer to this as the \textit{Inverse Floquet Transform}. One can see that a super-cell of $N$ copies leads to effective periodicity $a' = Na$ in the calculations above. This is equivalent to $N$ equidistant sampling points spread over the Brillouin zone. In the following we will thus use the super-cell size $N$ to denote the sampling rate of the Brillouin zone. Halving the sample rate thus effectively doubles the size of the super-cell.

\subsection{Numerical example: Near-field of a focused Gaussian beam in a periodic line grating}
We apply the  Inverse Floquet Transform to a numerical example. We aim to model the near-field of a focused Gaussian beam in a periodic line grating. As a model for the grating we use a setup which has been described also previously~\cite{SchneiderSantiagoSoltwischetal.2019,SchneiderHammerschmidtZschiedrichetal.2019,FarchminHammerschmidtSchneideretal.2019}. Here, the silicon grating is illuminated with a 2D Gaussian beam of the vacuum wavelength $\lambda_0 = 265\,$nm and of a waist of $2\lambda_0$. All evanescent waves are truncated by passing the beam through an optical system with an aperture of  $NA=1$. 

   \begin{figure} [ht]
   \begin{center}
   \begin{tabular}{c} 
   \includegraphics[height=4.5cm]{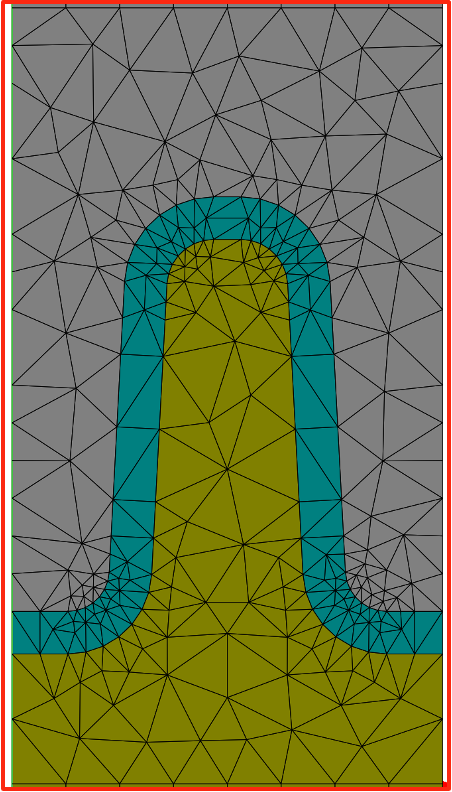}
   \includegraphics[height=5cm]{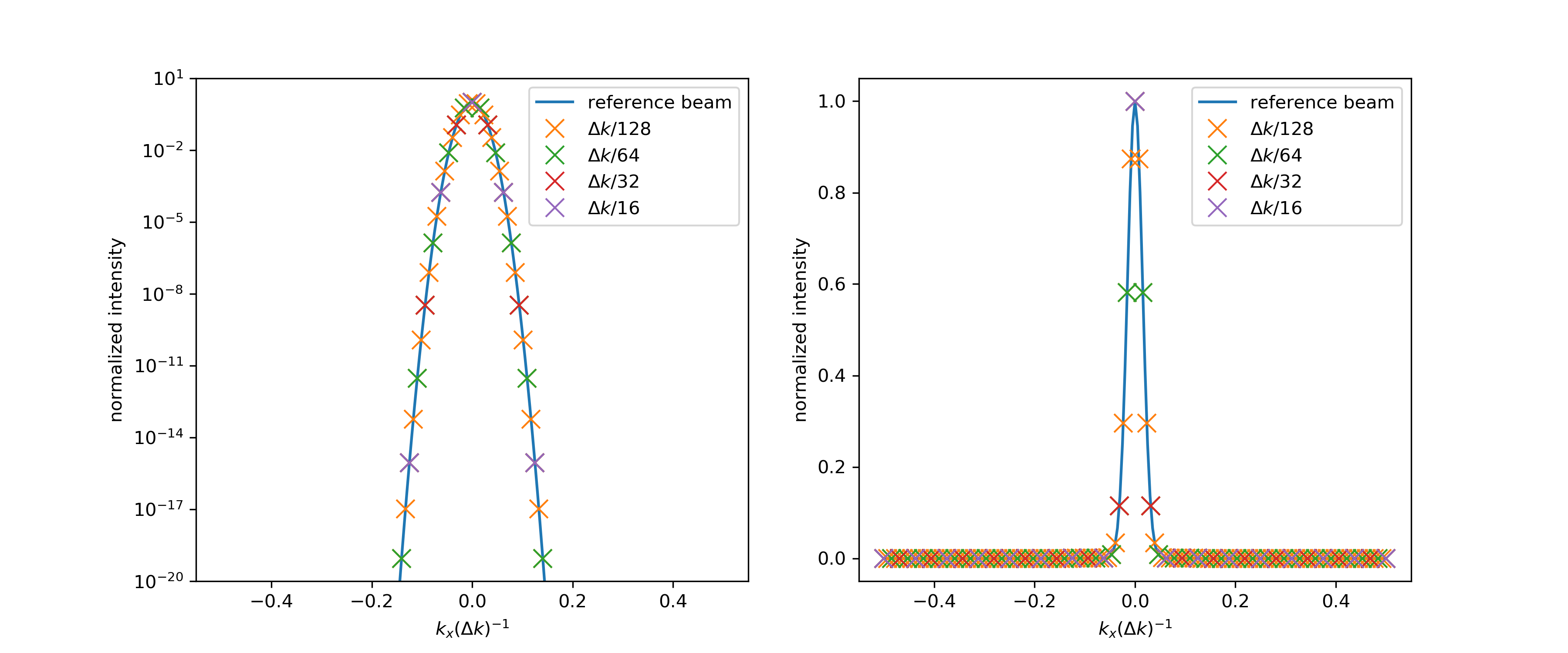}
   \end{tabular}
   \end{center}
   \caption[beamsampling] 
   { \label{fig:beamsampling} 
\textit{Left:} FEM-mesh of the investigate line grating with corner rounding and a conformal oxide layer above the Si-line. \textit{Center:} Normalized intensity of the Gaussian beam over the Brillouin zone for different equidistant sampling rates on a log-scale (center) and linear-scale (right). Due to the small spectral width of the beam a reasonable sampling rate is required for an accurate representation.}
   \end{figure} 
   
In Figure \ref{fig:beamsampling} the geometry of the line is shown together with the intensity plots of the Gaussian beam over the Brillouin zone on a log and linear scale. The  Brillouin zone is plotted from $-0.5$ to $0.5$ to better visualize the familiar symmetry of the Gaussian. Here we use $\Delta k = \frac{2\pi}{50\text{nm}}$ for the $x$ axis. The linear scale best demonstrates the limited bandwidth of the beam in k-space. The sampling positions corresponding to different super-cells are indicated by coloured crosses. Due to the small width it is immediately obvious that a $N=16$ is insufficient to sample the beam as we have only the $\mathbf{k_B}=0$ point with a meaningful weight in this expansion. 

We calculate the Floquet transform of the beam and use it as source fields for our FEM calculations on the unit cells. We employ the commercial solver \JCMsuite to solve these scattering problems. It is obvious from Figure \ref{fig:beamsampling} that we can neglect $\vert\mathbf{k_B}\vert>0.1$ without any loss of accuracy. Halving the sample rate effectively doubles the size of the super-cell and allows to re-use calculations when investigating the convergence.

   \begin{figure} [ht]
   \begin{center}
   \begin{tabular}{c} 
   \includegraphics[width=.9\linewidth]{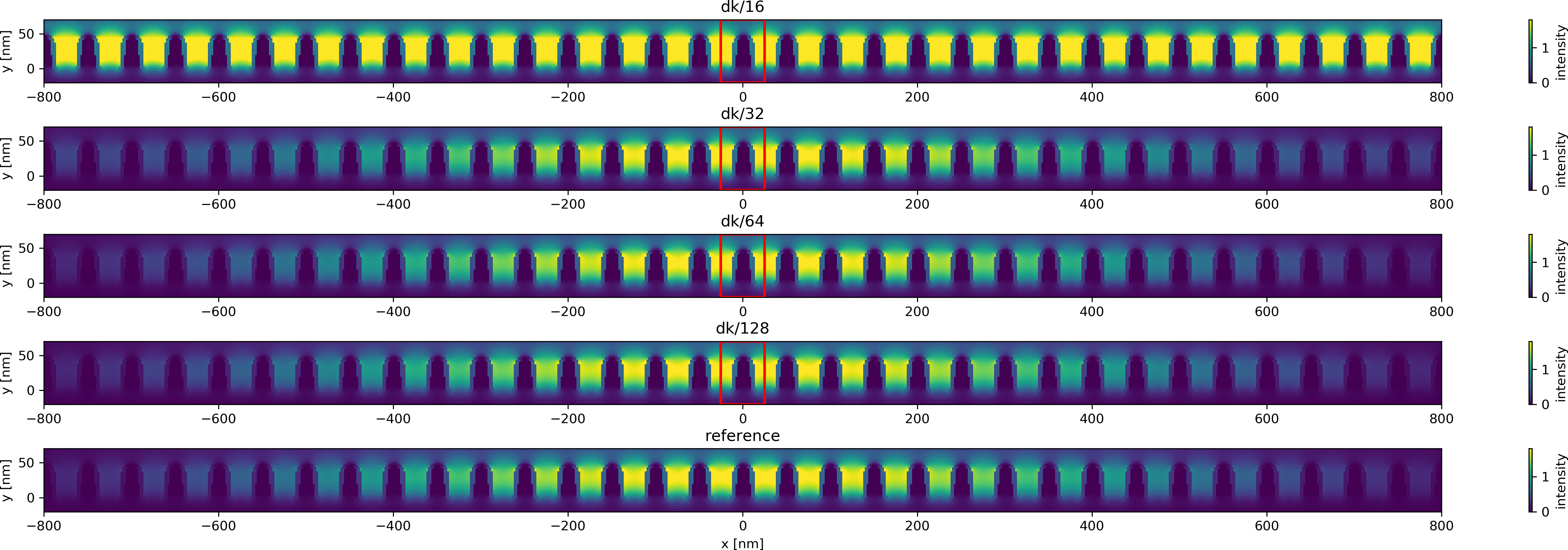}
   \end{tabular}
   \end{center}
   \caption[supercells] 
   { \label{fig:overview_supercells} 
    Exports of the electric field intensity in a cutout of the periodic grating around the focus of the incident beam. The unit cell used for computations is highlighted in red. The sampling rate in k-space is indicated above each plot. The bottom plot shows the reference calculation with a finite grating width. While the sampling rate of dk/16 is too coarse to form a beam, even at dk/32 the intensity pattern is visually almost indistinguishable from the dk/128 or the reference data set.  
    }   
   \end{figure} 
   \begin{figure} [ht]
   \begin{center}
   \begin{tabular}{c} 
   \includegraphics[height=5cm]{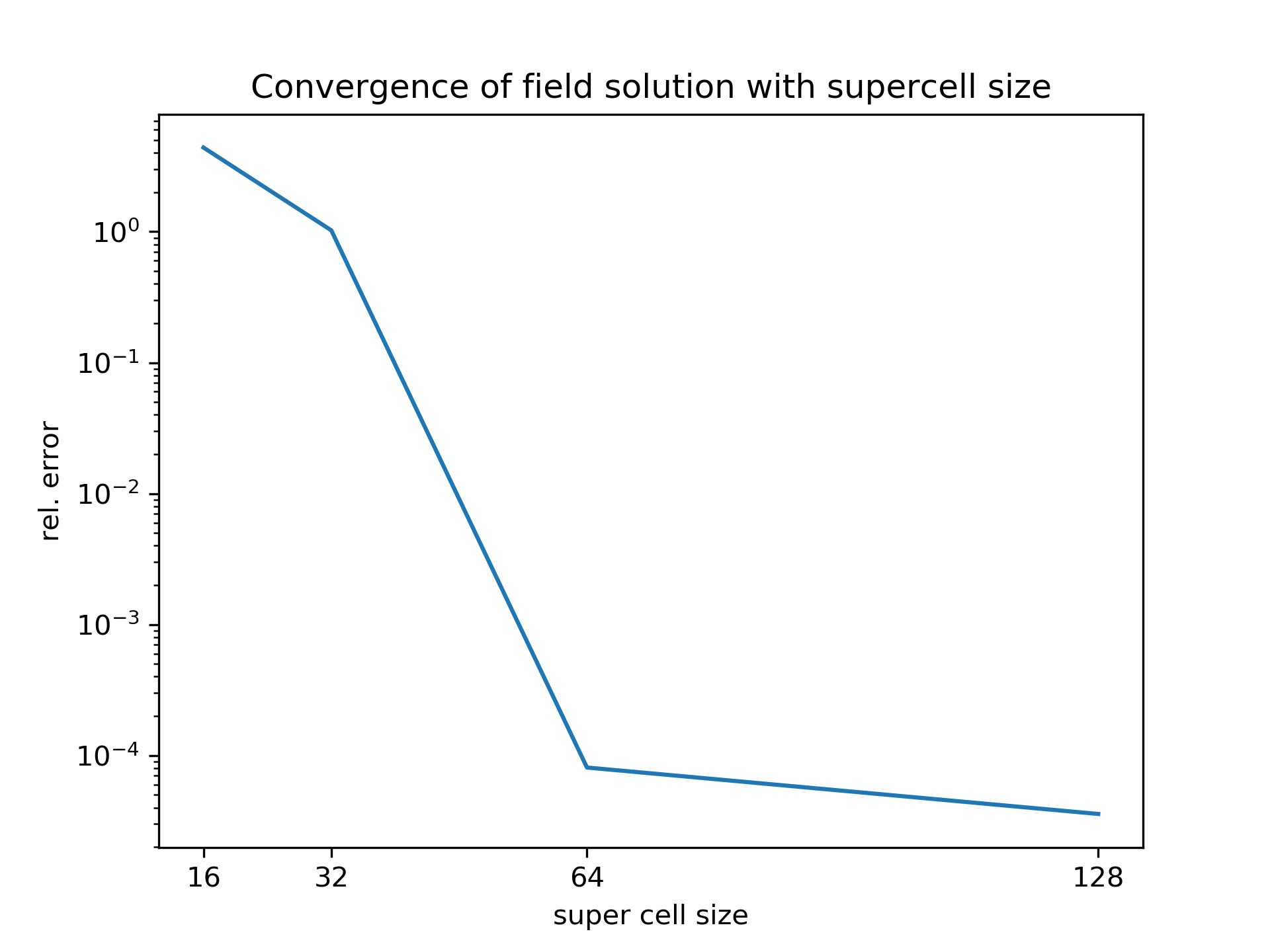}
   \end{tabular}
   \end{center}
   \caption[pupilpattern] 
   { \label{fig:convergence} 
Relative error in the electric fields compared to the reference simulation for increasing sampling rate measured of a super-cell of size 128. The sampling rate is represented as the corresponding super-cell size. One observes convergence with the number of super-cells. The apparent drop of in the convergence rate beyond 64 can be attributed to the comparison with the finite size aperiodic reference.
}
   \end{figure} 

Figure \ref{fig:overview_supercells} shows exports of the electric field intensity in a cutout of the periodic grating around the focus of the incident beam. The unit cell used for computations is highlighted in red. The sampling rate in $k$-space is indicated above each plot. The bottom plot shows the reference calculation with a finite grating width. Above we have already found the sampling rate of $N = 16$ too coarse to sample the Gaussian beam. This is apparent from these exports where the field is not localized. At $N=32$ the intensity pattern is visually almost indistinguishable from the $N=128$ or the reference data set. Here we have chosen a finite sized grating with 128 unit cells illuminated with the same beam as the aperiodic reference. 

In contrast to the qualitative comparison in figure \ref{fig:overview_supercells} we show the convergence of the relative error in the fields in the $L^2$ norm with increasing super-cell size in Figure \ref{fig:convergence}. We observe a drastic reduction in the error by 4 orders of magnitude moving from $N=32$ to $N=64$. The apparent drop of in the convergence rate beyond $N=64$ can be attributed to the comparison with the finite size aperiodic reference. 

\section{Forward Modelling for Coherent Fourier Scatterometry}

\label{sec:imaging}
Calculating the far-field from a near-field reconstructed with the Inverse Floquet Transform is unnecessarily cumbersome as when can exploit the linearity of the Fourier Transform and directly work on the Fourier transforms of the individual calculations of the Floquet Transforms. 
In this section we describe the workflow of the suggested forward model for coherent beams to be used for Coherent Fourier Scatterometry. Subsequently we employ the method to model the far-field scattering of a coherent focused beam by a line grating measured which is measured  in the back focal plane of a complex optical system.
\subsection{Fourier-optics based forward model}
The proposed forward model employs three steps to compute a forward pass in the coherent Fourier scatterometry. 
\begin{enumerate}
    \item pre-compute a library of Fourier transforms in reflection for plane wave illuminations covering the input pupil
    \item modelling of an incident beam in k-space
    \item compute propagation of the beam through the optical system
\end{enumerate}

The first step (building a pre-computed library of scattering simulations with plane wave illuminations of the investigated scatters) may take seconds to minutes or even longer for large 3D computational domains, high discretisation accuracies and high sampling densities of the illumination pupil. The subsequent steps require only the Fourier transforms of the reflected fields and not the near-field solutions themselves. The employed FEM solver \JCMsuite computes the solutions for all incident directions (and polarisations ) in a Bloch-family in one FEM simulation with little overhead. We consider this step computationally expensive but this is essentially an offline step that is embarrassingly parallel.

The second step requires a description of the coherent incident beam in the Fourier domain. Typically illuminations are described as circular, annular, or Gaussian distributions in k-space. But more complex structured multi-pole or Quasar illuminations are easy to describe the in same frame work. All illuminations are characterised a both (in-) coherent or partially coherent beams via their intensity, polarisation and phase distribution over the incident pupil. We have developed input sampling generators with adjustable accuracy for all the illuminations listed above.  

The final step is to compute the propagation of the field excited by a given point in the illumination pupil through an optical system. This is computed in a Fourier optics framework in the `OpticalImaging` tool chain of \JCMsuite. In this framework optical aberrations and misalignment or shifts of the optical systems can be considered as well as restrictions due to apertures.

The last two steps are fairly light-weight compared to the computationally more expensive first step as they only involve Fourier optics. 

The coherent superposition of all the fields yields an approximation of the solution in a continuous Fourier representation. The power spectral density of this field corresponds to the image observed in the back focal plane of the imaging system. 
We employ Hopkins’ interpolation of the data points within the library to model the reflection of beam sampling points not exactly represented in the data. The resulting coherent continuous Fourier representation is up-sampled to the desired output resolution. 

A periodic scattering target yields discrete diffraction orders in the Fourier transform that are not necessarily compatible with a given output resolution. Interpolation of these on a non-matching output sampling is possible, but it leads to noisy artifacts in the power spectral densities while the real-space images are indistinguishable on typical scales. As both the sampling of the output Fourier transform and the illumination sampling are arbitrary they can be chosen to match the reciprocal lattice of the periodic sample, thus avoiding sampling artifacts.

\subsection{Numerical example: scattering of a coherent focused beam by a line grating}
\label{sec:cfs_example}
In the following we consider the scattering of a coherent focused beam by a silicon line grating on a silicon substrate into air. The line is modelled with a CD of $294\,$nm, a height of $150\,$nm with a SW of $90\deg$ at a period of $589\,$nm. At the illumination wavelength of $488\,$nm we use $n_{Si}=4.3682+0.07971$ as the refractive index of silicon. The incident beam is modelled with a constant amplitude and phase over the illumination pupil with $ NA = 0.8$. 

\begin{figure} [ht]
   \begin{center}
   \begin{tabular}{c} 
   \includegraphics[height=5cm]{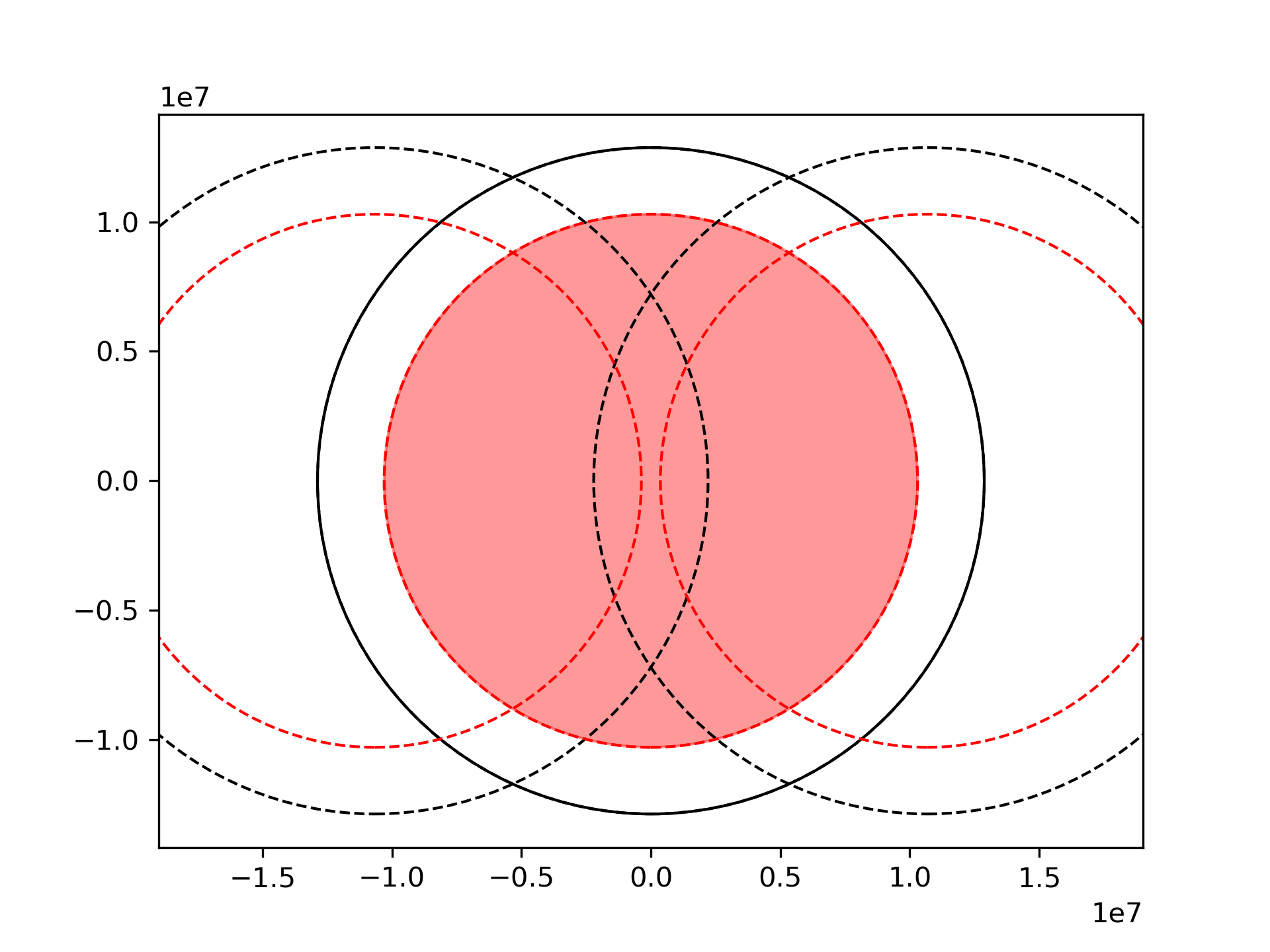}
   \includegraphics[height=5cm]{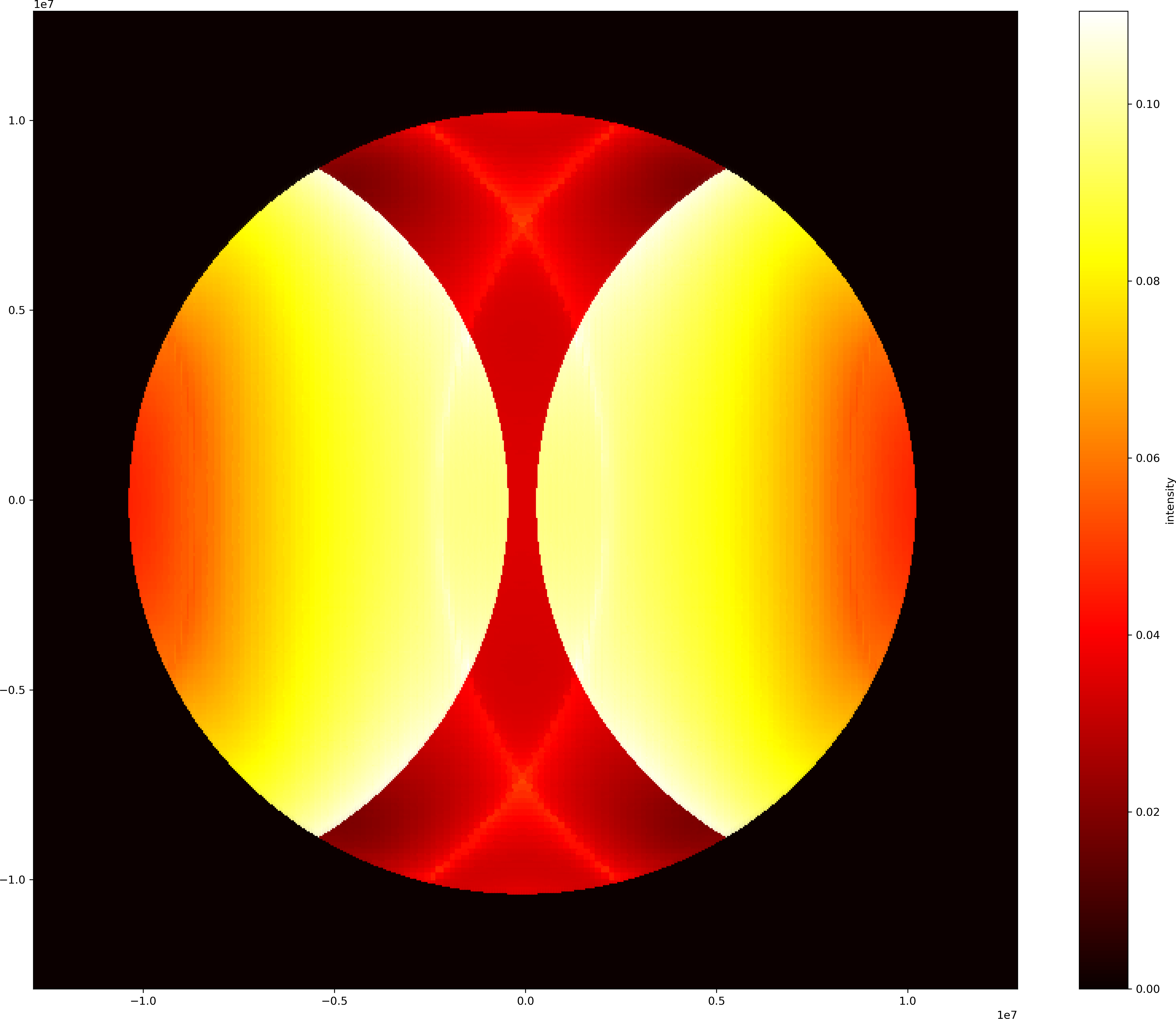} 
   \end{tabular}
   \end{center}
   \caption[pupilpattern] 
   { \label{fig:pupil_pattern} 
Idealized and modelled pupil patterns for the coherent Fourier scatterometry setup. \textit{Left:} Light in the red part of the pupil with NA=0.8 (black solid line NA=1.0) can be captured by the objective. The cutoff of light diffracted light by the NA of the optical system (red-dashed) or the propagating spectrum (black-dashed) provide structure in the pupil image. \textit{Right:} Simulated pupil pattern for the line grating. The features outlined on the left are clearly observable.

    } 

   \end{figure}

The scattered spectrum of the beam yields discrete diffraction orders in the $k_x$ direction (where we assume the grating is periodic in the x-coordinate and in variant along y) and is continuous in $k_y$. We observe the $\pm 1$ diffraction orders within the output pupil of the optical system in addition to the $0-th$ for this combination of wavelength and periodicity. Figure \ref{fig:pupil_pattern} shows an idealized and the computed intensity pattern over the pupil. The characteristic circles due to the disappearance of diffraction orders from either the propagating spectrum (dashed black circles) or the collection NA (dashed red circles) are clearly visible and well resolved by the simulation.

\subsubsection{Influence of the library size}

\begin{figure} [ht]
   \begin{center}
   \begin{tabular}{c} 
   \includegraphics[height=5cm]{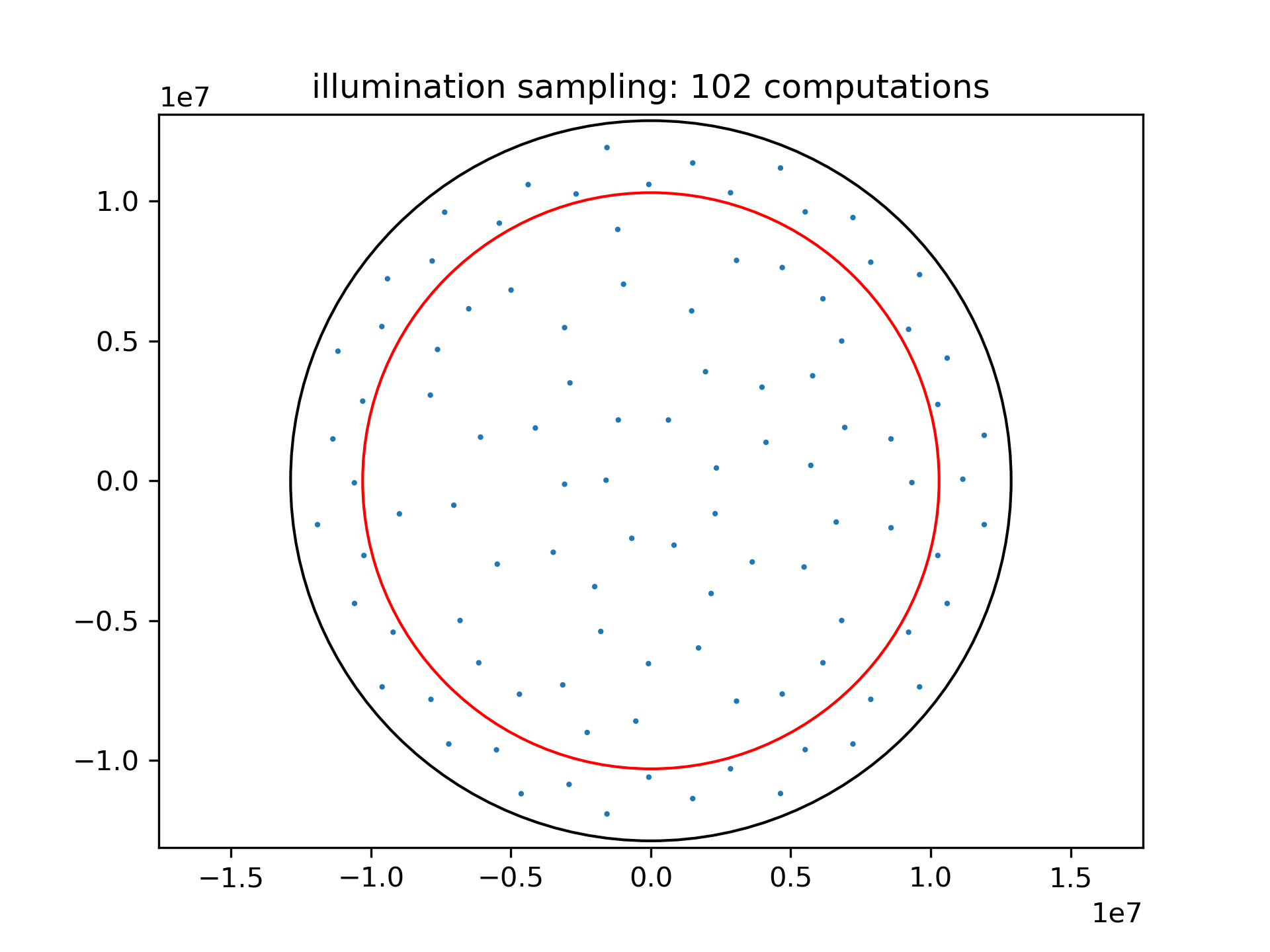} 
   \includegraphics[height=5cm]{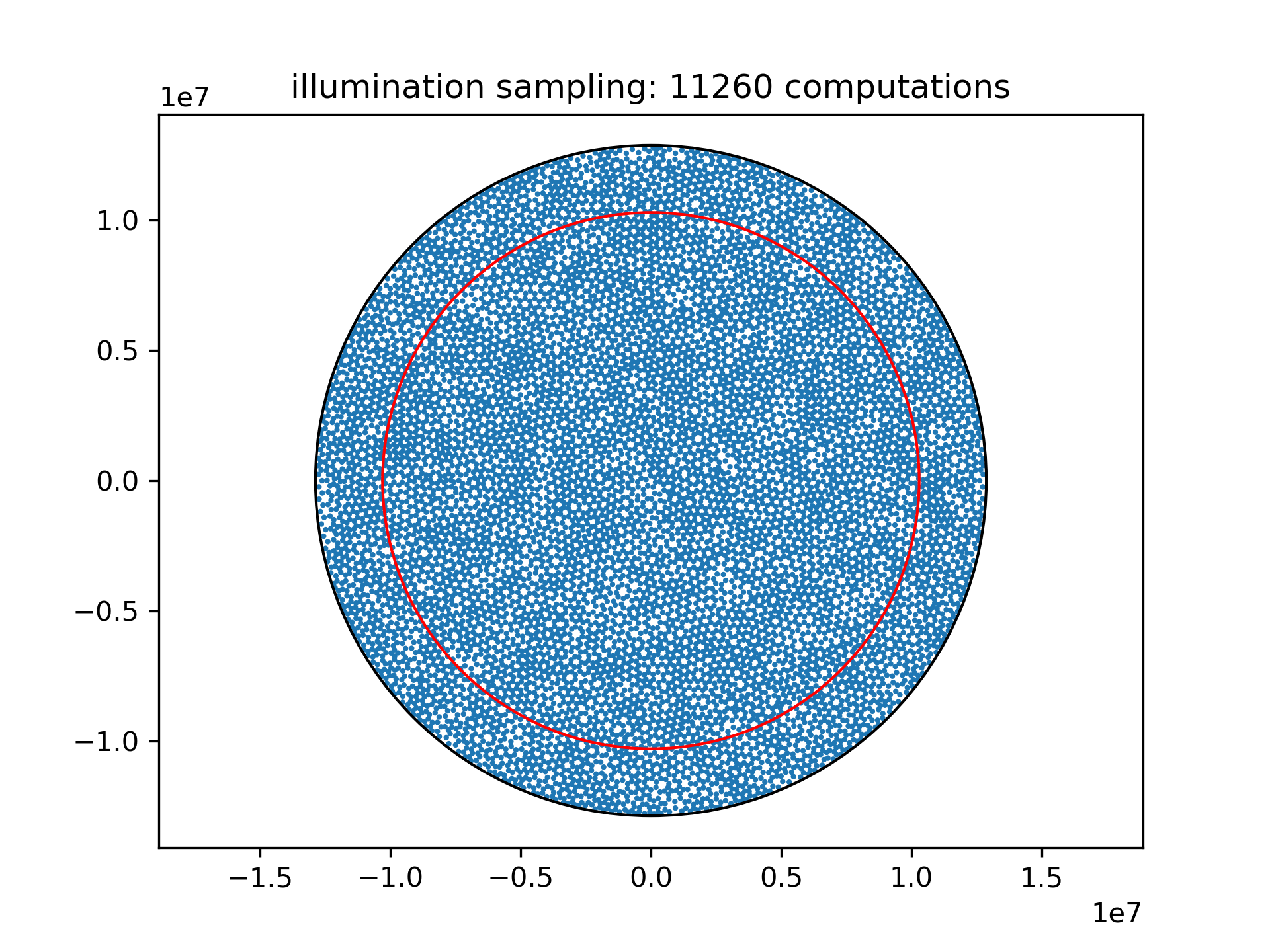} \\ 
   \includegraphics[height=5cm]{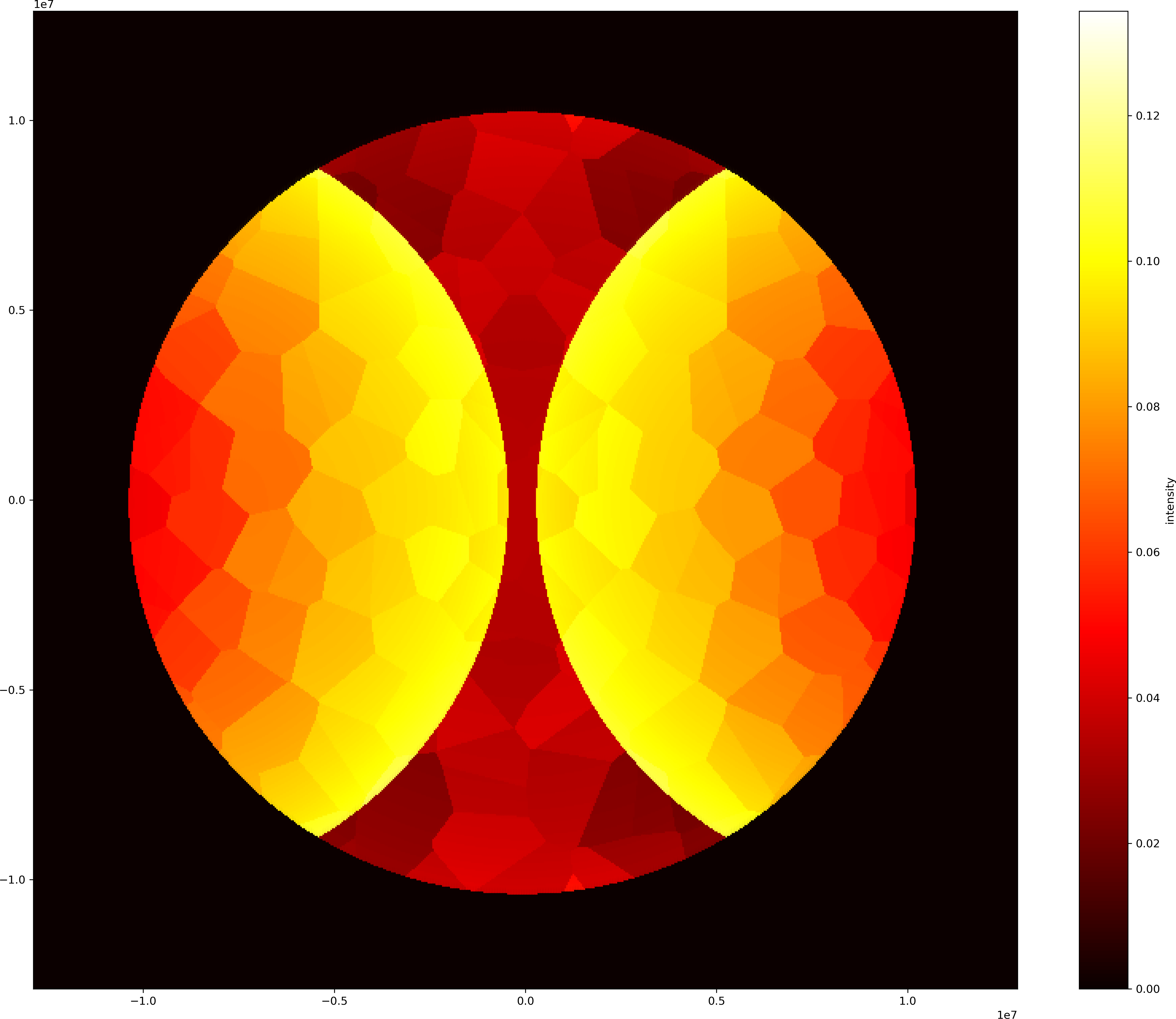} \hspace{.9cm}
   \includegraphics[height=5cm]{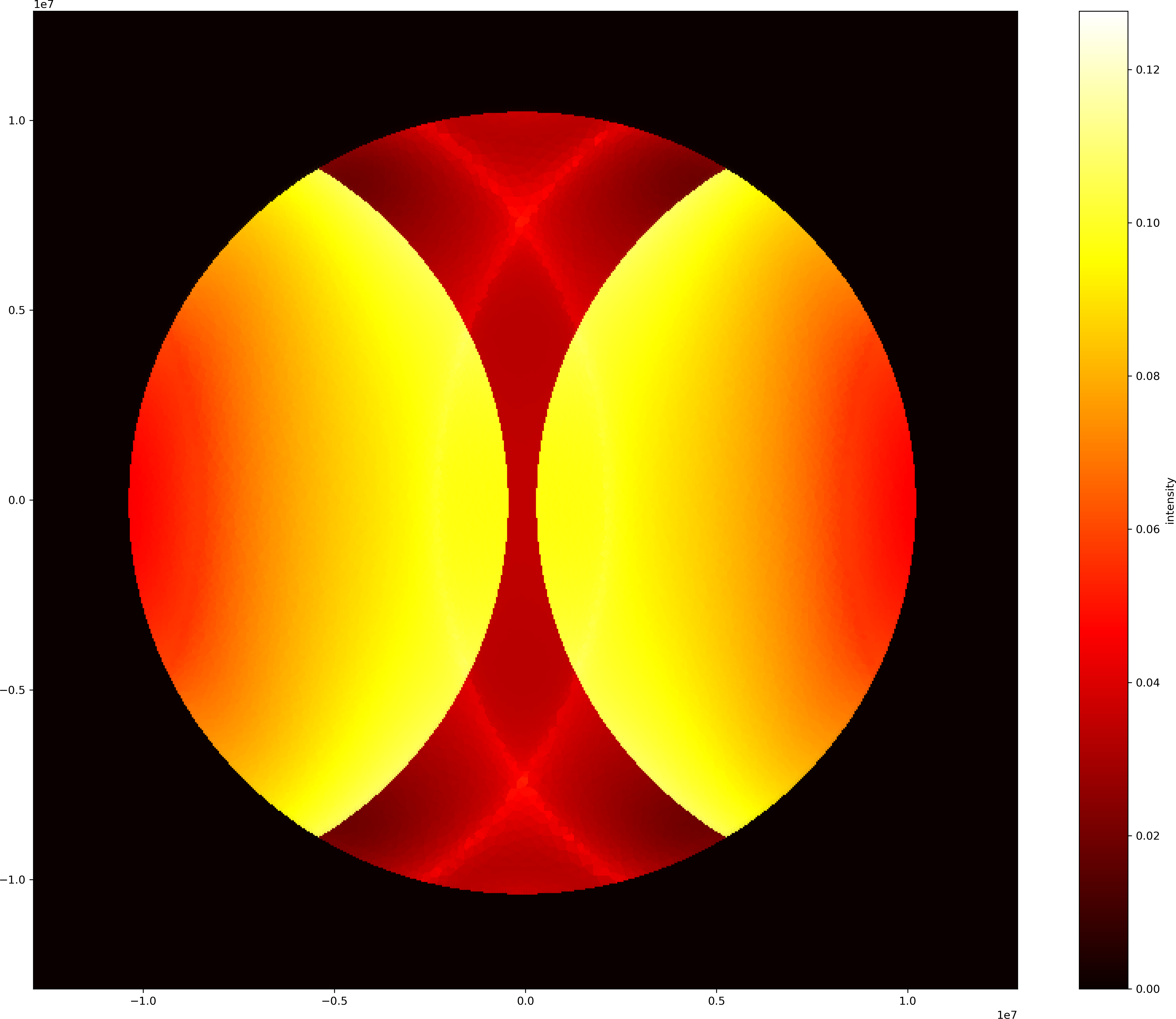} \\ 
   \end{tabular}
   \end{center}
   \caption[illuminationSampling] 
   { \label{fig:pupil_sampling} \textit{Top:}
Location of the pre-calculated illumination points in the pupil for two different dimensions of the offline database. Points outside the collection NA (red circle) are considered as well due to their Bloch families. 
\textit{Bottom:}
Pupil images for fine illumination pupil and the output sampling. Interpolation artifacts are due to the nearest neighbor interpolation. 
}
\end{figure} 
   
The presented method employs Hopkins’ interpolation of the pre-computed data points within the library. This of course leads to interpolation artifacts in the resulting pupil images. We thus compare two extreme cases of the unstructured sampling of the illumination pupil: (a) 102 and (b) 11260 reference FEM computations. Both samplings are depicted in the upper row in Figure \ref{fig:pupil_sampling}. 

We chose a very fine sampling for both the illumination pupil and the output sampling of $512\times512$ points over the pupil. The resulting pupil plots are shown in the bottom row in  Figure \ref{fig:pupil_sampling}. Interpolation artifacts due to the nearest neighbor interpolation are visible, but inconsequential for the high-resolution library. The low-resolution image yields clearly visible Voronoi-patches due to the interpolation. Nevertheless, the intensity distribution is well approximated while the fine structure of the critical circles cannot be reproduced as it is missing from the data.

\subsubsection{Efficiency considerations}
In the considerations above we have worked with an unstructured sampling of the illumination pupil. This is of course not the most efficient sampling technique. In Section \ref{sec:invFloquetTrafo} we exploited the fact that we can reduce the computational effort to the first Brillouin zone. Exploiting symmetries there are further reductions possible. 

Solving the scattering problem with Bloch periodic boundary conditions for a plane wave with transverse wave vector $\mathbf{k_{\perp}} = \mathbf{k_B}$ for some $ \mathbf{k_B}$ we immediately find the same boundary condition for all $\mathbf{k_{\perp}} \in \{ \mathbf{k_B}+\mathbf{b}m \vert m \in \bbz\}$. We call this set the Bloch family of $ \mathbf{k_B}$. \JCMsuite allows to calculate all these incident fields simultaneously with very little overhead. The same holds of course for different polarisations. This can further reduce the required effort significantly for setups with larger periods where more members of the  Bloch family correspond to propagating plane waves.

\section{Conclusion}
\label{sec:conclusion}
We have presented two modelling techniques for scattering of coherent focused beams by periodic structures. The Inverse Floquet Transform was applied to model the near-fields in a periodic grating arrangement when illuminated with a focused Gaussian beam. A second model, tailored for applications in the far-field regime, was presented and applied for a line grating to be measured with Coherent Fourier Scatterometry. The Fourier optics based model allows to include optical aberrations in optical systems used for both the illumination and detection of the scattered fields.

\appendix    

\acknowledgments 
 This project has received funding from the EMPIR programme co-financed by the Participating States and from the European Union’s Horizon 2020 research and innovation programme (project 20FUN02 "POLIGHT"). This project has received funding from the EMPIR programme co-financed by the Participating States and from the European Union’s Horizon 2020 research and innovation programme (project 20IND04 "ATMOC"). This project is funded by the German Federal Ministry of Education and Research (BMBF, project number 01IS20080A, SiM4diM). 

\bibliography{main} 
\bibliographystyle{spiebib} 

\end{document}